\documentclass[aps,prl,twocolumn,superscriptaddress]{revtex4-1}
\usepackage{graphicx}
\usepackage{dcolumn}
\usepackage{bm}
\usepackage{color}
\usepackage{ulem}
\usepackage[usenames,dvipsnames,svgnames,table]{xcolor}

\newcommand{\pstar}{$p^{\star}$}

\newcommand{\pcdw}{$p_{\rm CDW}$}

\newcommand{\Tc}{$T_{\rm c}$}
\newcommand{\Tstar}{$T^{\star}$}

\newcommand{\Tcdw}{$T_{\rm CDW}$}
\newcommand{\Tmax}{$T_{\rm max}$}

\begin{document}



\title{Effect of pressure on the pseudogap and charge-density-wave phases of the cuprate Nd-LSCO probed by thermopower measurements}

\author{A.~Gourgout}
\affiliation{Institut Quantique, D\'epartement de physique \& RQMP, Universit\'e de Sherbrooke, Sherbrooke, Qu\'ebec J1K 2R1, Canada}

\author{A.~Ataei}
\affiliation{Institut Quantique, D\'epartement de physique \& RQMP, Universit\'e de Sherbrooke, Sherbrooke, Qu\'ebec J1K 2R1, Canada}

\author{M.-E.~Boulanger}
\affiliation{Institut Quantique, D\'epartement de physique \& RQMP, Universit\'e de Sherbrooke, Sherbrooke, Qu\'ebec J1K 2R1, Canada}

\author{S.~Badoux}
\affiliation{Institut Quantique, D\'epartement de physique \& RQMP, Universit\'e de Sherbrooke, Sherbrooke, Qu\'ebec J1K 2R1, Canada}

\author{S.~Th\'eriault}
\affiliation{Institut Quantique, D\'epartement de physique \& RQMP, Universit\'e de Sherbrooke, Sherbrooke, Qu\'ebec J1K 2R1, Canada}

\author{D.~Graf}
\affiliation{National High Magnetic Field Laboratory, Florida State University, Tallahassee, FL 32306, USA}

\author{J.-S.~Zhou}
\affiliation{Texas Materials Institute, University of Texas - Austin, Austin, Texas 78712, USA}

\author{S.~Pyon}
\affiliation{Department of Advanced Materials Science, University of Tokyo, Kashiwa, Japan}
\affiliation{Department of Applied Physics, University of Tokyo, Tokyo, Japan}

\author{T.~Takayama}
\affiliation{Department of Advanced Materials Science, University of Tokyo, Kashiwa, Japan}
\affiliation{Max Planck Institute for Solid State Research, Stuttgart, Germany}

\author{H.~Takagi}
\affiliation{Department of Advanced Materials Science, University of Tokyo, Kashiwa, Japan}
\affiliation{Max Planck Institute for Solid State Research, Stuttgart, Germany}
\affiliation{Department of Physics, University of Tokyo, Tokyo, Japan}
\affiliation{Institute for Functional Matter and Quantum Technologies, University of Stuttgart, Stuttgart, Germany}

\author{Nicolas~Doiron-Leyraud}
\email[]{nicolas.doiron-leyraud@usherbrooke.ca}
\affiliation{Institut Quantique, D\'epartement de physique \& RQMP, Universit\'e de Sherbrooke, Sherbrooke, Qu\'ebec J1K 2R1, Canada}

\author{Louis~Taillefer}
\email[]{louis.taillefer@usherbrooke.ca}
\affiliation{Institut Quantique, D\'epartement de physique \& RQMP, Universit\'e de Sherbrooke, Sherbrooke, Qu\'ebec J1K 2R1, Canada}
\affiliation{Canadian Institute for Advanced Research, Toronto, Ontario M5G 1M1, Canada}

\date{\today}

\begin{abstract}

We report thermopower measurements under hydrostatic pressure on the cuprate superconductor La$_{1.6-x}$Nd$_{0.4}$Sr$_x$CuO$_4$ (Nd-LSCO), at low-temperature in the normal state accessed by suppressing superconductivity with a magnetic field up to $H = 31$~T.
Using a newly developed AC thermopower measurement technique suitable for high pressure and high field, we track the pressure evolution of the Seebeck coefficient $S$.
At ambient pressure and low temperature, $S/T$ was recently found to suddenly increase in Nd-LSCO at the pseudogap critical doping $p^{\star} = 0.23$, consistent with a drop in carrier density $n$ from $n = 1 + p$ above $p^{\star}$ to $n = p$ below.
Under a pressure of 2.0~GPa, we observe that this jump in $S/T$ is suppressed.
This confirms a previous pressure study based on electrical resistivity and Hall effect which found $dp^{\star}/dP \simeq - 0.01$~holes/GPa, thereby reinforcing the interpretation that this effect is driven by the pressure-induced shift of the van Hove point.
It implies that the pseudogap only exists when the Fermi surface is hole-like, which puts strong constraints on theories of the pseudogap phase.
We also report thermopower measurements on Nd-LSCO and La$_{1.8-x}$Eu$_{0.2}$Sr$_x$CuO$_4$ in the charge density-wave phase near $p \sim 1/8$, which reveals a weakening of this phase under pressure.

\end{abstract}

\pacs{}

\maketitle


\section{Introduction}
The pseudogap phase of cuprates is arguably one of their chief mysteries.
Understanding the pseudogap phase of cuprates has been hindered in part by the fact that it does not exhibit a clear and well-defined symmetry-breaking phase transition at its characteristic temperature $T^{\star}$.
However, upon crossing the pseudogap end point at doping $p^{\star}$ - where $T^{\star}$ vanishes - at low temperatures, recent measurements have revealed a clear and abrupt change in carrier density $n$, going from $n = 1 + p$ above $p^{\star}$ to $n = p$ below.
These are based on high-field and low-temperature measurements of the Hall effect on YBa$_2$Cu$_3$O$_y$ (YBCO)~\cite{badoux2016} and La$_{1.6-x}$Nd$_{0.4}$Sr$_x$CuO$_4$ (Nd-LSCO)~\cite{collignon2017}, resistivity on La$_{2-x}$Sr$_x$CuO$_4$ (LSCO)~\cite{laliberte2016} and Nd-LSCO~\cite{collignon2017}, and thermal conductivity on Nd-LSCO~\cite{michon2018} and LSCO~\cite{bourgeois-hope2019}.
Recently, the single-layer cuprates Bi$_2$Sr$_{2-x}$La$_x$CuO$_{6+\delta}$ (Bi2201)~\cite{lizaire2020,putzke2020} and Tl$_2$Ba$_2$CuO$_{6+\delta}$ (Tl2201)~\cite{putzke2020} were also shown to display such a drop in carrier density at $p^{\star}$ via Hall effect measurements, lending a sense of universality to these transport signatures of the pseudogap.
Thermodynamic measurements have recently revealed a logarithmic divergence of the electronic specific heat at $p^{\star}$~\cite{michon2019}, both as a function of doping and temperature, a classic signature of a quantum phase transition.
Above $p^{\star}$, the resistivity of Nd-LSCO~\cite{daou2009} and LSCO~\cite{cooper2009} exhibits a purely linear temperature dependence at low $T$, another signature of quantum criticality, with a slope that reaches the Planckian limit~\cite{legros2019}.

In the LSCO-based family of cuprates, an intriguing question is why do LSCO and Nd-LSCO have such a different $p^{\star}$, namely \pstar~$\simeq 0.18$ for LSCO~\cite{cooper2009,laliberte2016}, and $p^{\star} = 0.23$ for Nd-LSCO~\cite{collignon2017}.
This issue was examined via electrical resistivity and Hall effect measurements under hydrostatic pressure on Nd-LSCO~\cite{doiron-leyraud2017}, which found, using the drop in carrier density as a clear marker of $p^{\star}$, that $p^{\star}$ moves down with pressure $P$ at a rate of $\delta p^{\star}/\delta P \sim - 0.01$ hole/Cu atom per GPa.
This effect was shown to be driven by the pressure displacement of the van Hove point~\cite{doiron-leyraud2017}, where the Fermi surface changes from hole-like to electron-like, demonstrating that the pseudogap phase can only exist on a hole-like Fermi surface.
In other words, the doping at which this Fermi surface change occurs, $p_{\rm FS}$, constrains the pseudogap phase, such that $p^{\star} \leq p_{\rm FS}$.
In Nd-LSCO, $p_{\rm FS} = 0.22 \pm 0.01$~\cite{matt2015} and $p^{\star} = 0.23 \pm 0.01$~\cite{collignon2017}; in LSCO, $p_{\rm FS} = 0.19 \pm 0.02$~\cite{yoshida2009,chang2008b} and \pstar~$\simeq 0.18$~\cite{cooper2009,laliberte2016}; in Bi2201, $p_{\rm FS} = 0.41 \pm 0.02$~\cite{kondo2004} and $p^{\star} = 0.40 \pm 0.01$~\cite{lizaire2020}.
This explains why $p^{\star}$ is different in these cuprates.
The fact that $p^{\star} \leq p_{\rm FS}$ places a strong constraint on candidate theories of the pseudogap phase, and it was found to be consistent with numerical solutions of the Hubbard model~\cite{wu2018}.

Recently, the thermopower was used as a probe of the carrier density across $p^{\star}$ in Nd-LSCO~\cite{collignon2020}.
Unlike the Hall coefficient, $S$ does not depend sensitively on the curvature or shape of the Fermi surface.
Unlike the conductivity (electrical or thermal), it does not depend sensitively on the level of impurity scattering.
Within a simple model, in the $T = 0$ limit and for a single band, the Seebeck coefficient of thermopower depends on two parameters only, the effective mass $m^{\star}$ and the carrier density $n$~\cite{behnia2004,miyake2005}:
\begin{equation}
\frac{S}{T} \propto \frac{m^{\star}}{n}.
\end{equation}
Fundamentally, $S/T$ is the specific heat per carrier.
While seemingly over-simplistic, this expression was shown~\cite{behnia2004} to hold even in the presence of multiple bands and strong electronic correlations, as evidenced for a great variety of materials that includes common metals, oxides, heavy fermions, cuprates, and organic superconductors.
In YBCO at $p = 0.11$ for instance, there is excellent quantitative agreement between the normal-state $S/T$ measured in the $T \rightarrow 0$ limit~\cite{laliberte2011,doiron-leyraud2015} and the estimate from Eq.~1 using the carrier density $n$ and the effective mass $m^{\star}$ obtained from quantum oscillations~\cite{doiron-leyraud2007}.

\begin{figure}[t!]
\includegraphics[width=0.46\textwidth]{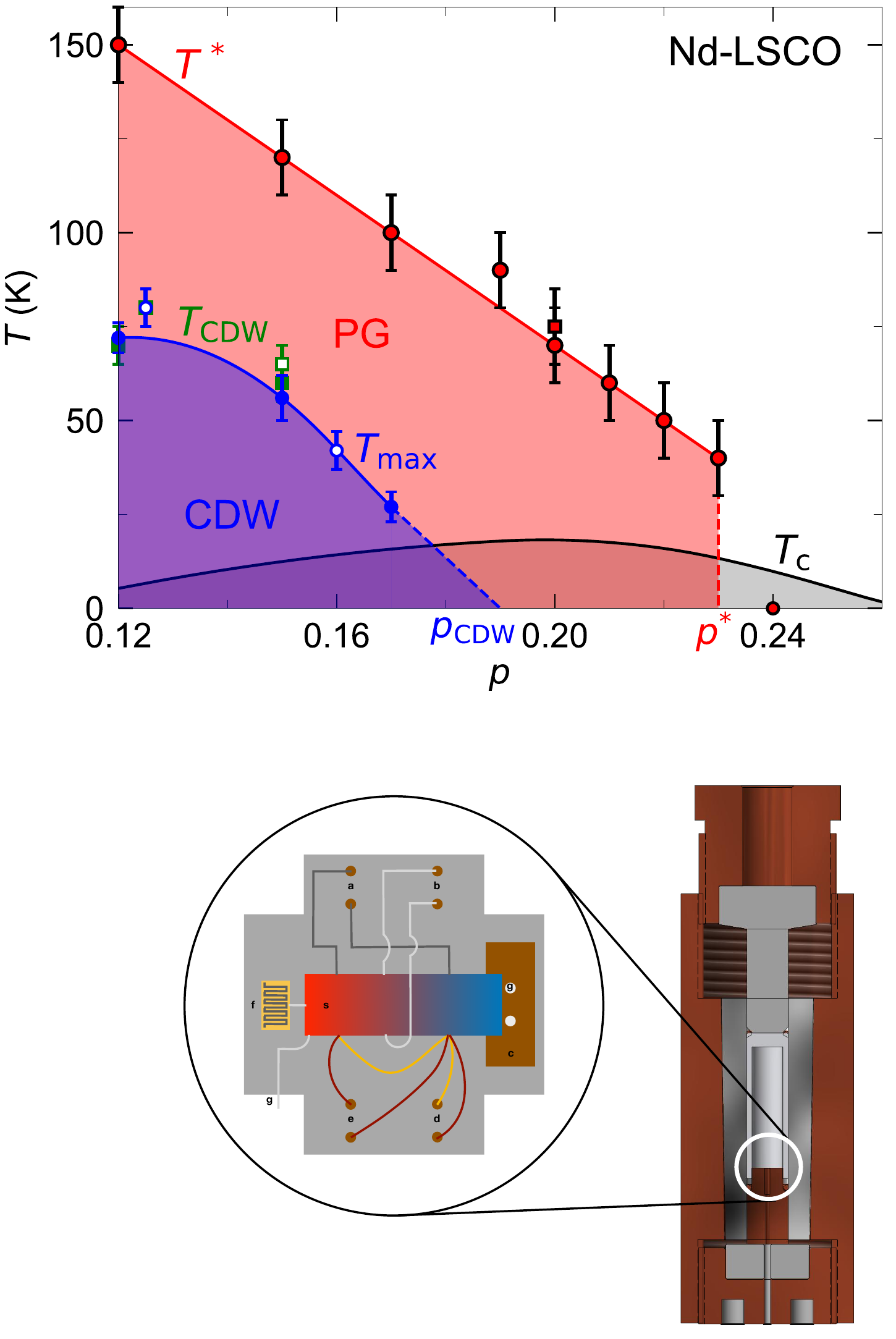}

\caption{
{\it Top:} Temperature-doping phase diagram of Nd-LSCO showing 
the pseudogap temperature \Tstar~extracted from resistivity (red dots~\cite{collignon2017,collignon2020}) 
and ARPES measurements (red squares, from ref.~\onlinecite{matt2015}), 
the CDW ordering temperature \Tcdw~as seen in x-ray diffraction measurements (green squares~\cite{zimmermann1998,niemoller1999}), 
the temperature \Tmax~of the maximum in $S/T$ vs $T$ (blue dots~\cite{collignon2020}), 
and a schematic of the zero-field superconducting transition temperature \Tc~(black line). 
Corresponding values for \Tmax~\cite{laliberte2011} and \Tcdw~\cite{fink2011} in 
La$_{1.8-x}$Eu$_{0.2}$Sr$_x$CuO$_4$ (Eu-LSCO) are shown as open symbols.
The red and blue full lines are guides to the eye.
The red dashed line marks the end of the pseudogap phase, at the critical doping \pstar~$= 0.23 \pm 0.01$~\cite{collignon2017}.
The blue dashed line is a linear extension of the full blue line, extrapolating to $p =$~\pcdw~$= 0.19 \pm 0.01$ at $T=0$.
{\it Bottom:} Illustration of our experimental setup, showing the piston-cylinder pressure cell and a zoom on the top view of our thermopower mount at the tip of the electrical feedthrough. The mount shows the sample (s; red to blue gradient), differential and absolute type E thermocouples (e,d; red), phosphor-bronze wires for $V_x$ pick-up (a; grey), strain gauge sample heater (f; yellow), and copper heat sink (c; brown). The entire setup is mounted on a small 2 mm x 2 mm G10 plate at the tip of our electrical feedthrough. Thermocouples are electrically insulated from the sample.
}
\label{Phase}
\end{figure}

In the present Article, we report our pressure study of the thermopower of Nd-LSCO and La$_{1.8-x}$Eu$_{0.2}$Sr$_x$CuO$_4$ (Eu-LSCO), single-layer, tetragonal cuprate superconductors with a low critical temperature $T_c$ and field $H_{c2}$, making them ideal candidates to study the field-induced normal-state Seebeck coefficient down to low temperatures.
The phase diagram of Nd-LSCO is shown in Fig.~\ref{Phase}, where the pseudogap temperature $T^{\star}$ extracted from resistivity measurements~\cite{collignon2020} is displayed and seen to be in agreement with angle-resolved photoemission spectroscopy (ARPES) measurements~\cite{matt2015}, showing that the transport signatures of the pseudogap in Nd-LSCO match those seen in spectroscopy.
In Nd-LSCO, recent thermopower measurements of the normal state found a sudden increase of $S/T$ below the pseudogap critical doping $p^{\star} = 0.23$.
This unambiguously confirms that the drop in $n$, from $n = 1 + p$ above $p^{\star}$ to $n = p$ below, first inferred from Hall effect, electrical resistivity, and thermal conductivity measurements is not an artifact of those transport properties but a genuine change in carrier density.
At lower doping, in the vicinity of $p \sim 1/8$, charge density-wave (CDW) order was previously shown to induce a negative $S/T$ at low temperature in YBCO~\cite{laliberte2011}, Eu-LSCO~\cite{laliberte2011}, HgBa$_2$CuO$_{4+\delta}$ (Hg1201)~\cite{doiron-leyraud2013}, and Nd-LSCO~\cite{collignon2020}.
The aim of the present study is to use the clear signatures of the pseudogap and CDW phases in thermopower to study their evolution with pressure.
To that effect, we recently developed a novel AC method which allows us to perform thermopower measurements under hydrostatic pressure and high magnetic field.

%
%

%
\begin{figure*}[t!]
\centering
\includegraphics[width=0.7\textwidth]{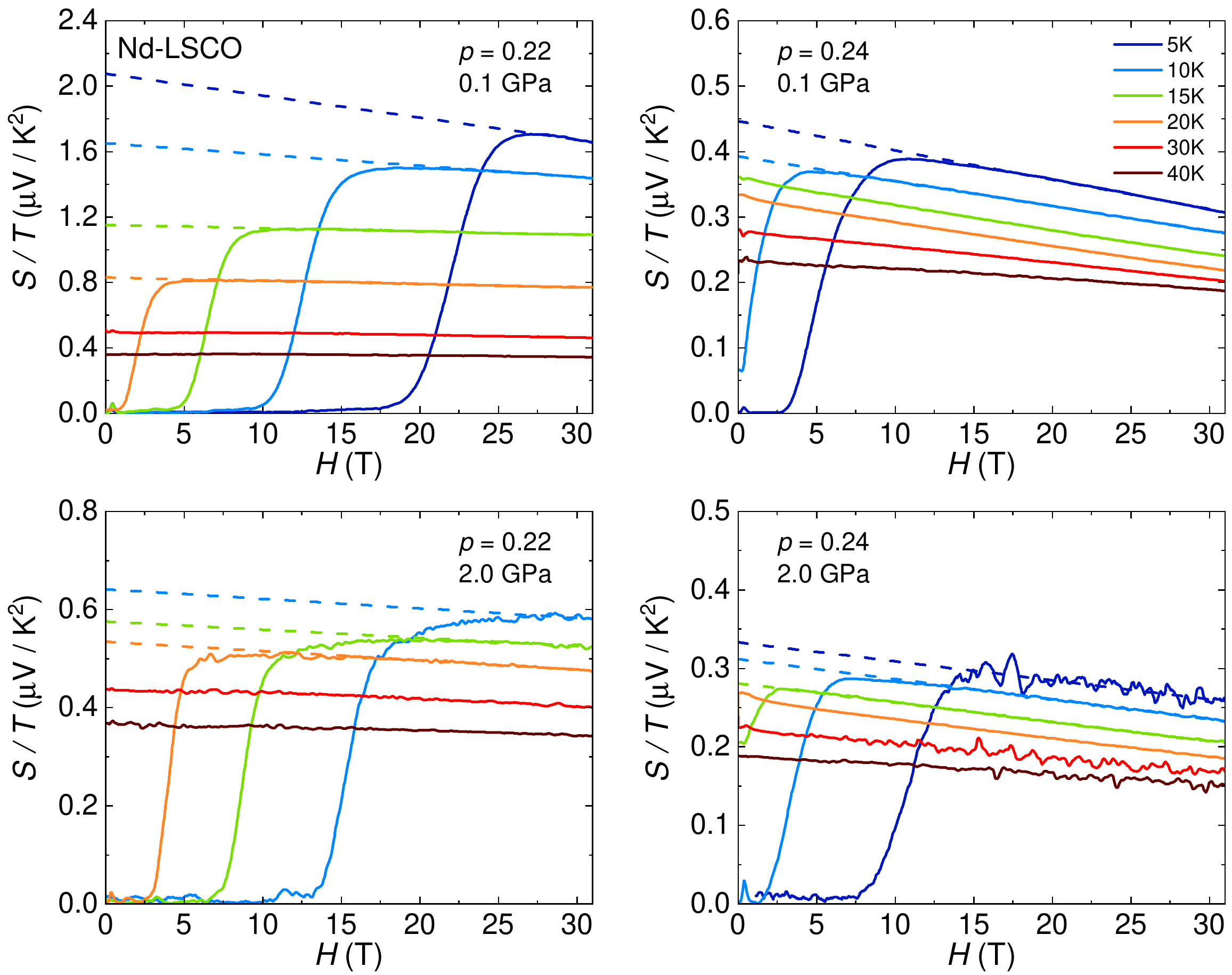}
\caption{
Isotherms of the Seebeck coefficient expressed as $S/T$ as a function of magnetic field $H$ in Nd-LSCO at $p = 0.22$ and 0.24, at temperature and pressure as indicated. As shown in the top panels, at near ambient pressure (0.1~GPa), the change of scale in $S/T$ by a factor of about 5 when going from $p = 0.22 < p^{\star}$ to $p = 0.24 > p^{\star}$ is readily visible in the raw data, indicative of the sudden drop in carrier density associated with the onset of the pseudogap phase~\cite{collignon2020}. Under 2.0~GPa, $S/T$ at $p = 0.22$ is heavily suppressed, signaling a lowering of $p^{\star}$ with pressure.
}
\label{SvsH}
\end{figure*}
%

Our main finding is a clear suppression of the pseudogap signature in thermopower in Nd-LSCO at $p = 0.22$, inside the pseudogap phase, while outside the pseudogap phase, at $p = 0.24$, the thermopower shows a marginal change with pressure.
This provides a clear and robust confirmation of the shift of $p^{\star}$ with pressure first deduced from electrical transport measurements~\cite{doiron-leyraud2017}.
In Nd-LSCO and Eu-LSCO in the CDW phase, at $p \sim 1/8$, we observe a suppression of both the negative amplitude of $S/T$ and its sign-change temperature with pressure, which we interpret as a weakening of CDW order with pressure, as also seen in YBCO via transport~\cite{cyr-choiniere2018}, x-rays~\cite{souliou2018}, and nuclear magnetic resonance (NMR)~\cite{vinograd2019} measurements.
We discuss the implications of our findings for the overall phase diagram of cuprates.
Our results highlight the probing power of thermopower measurements under pressure.

%
\begin{figure*}[t]
\centering
\includegraphics[width=0.9\textwidth]{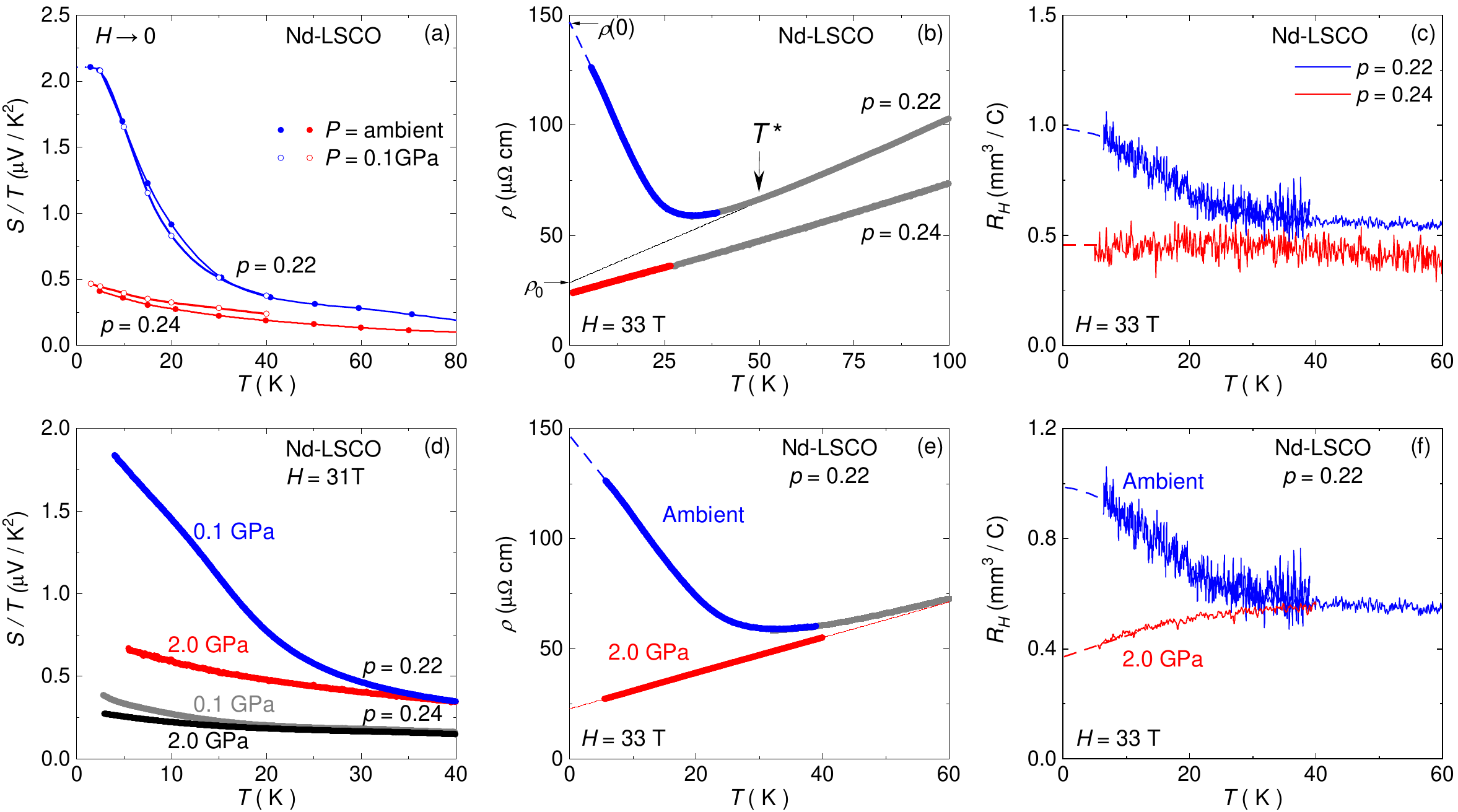}
\caption{Summary of pressure effects on the transport properties of Nd-LSCO near $p^{\star}$. Top row (ambient pressure): (a) Seebeck coefficient $S/T$, (b) electrical resistivity $\rho$, and (c) Hall coefficient $R_{\rm H}$ in Nd-LSCO at $p = 0.22$ and 0.24, in the field-induced normal-state. Data for $S/T$ are from the present study (0.1~GPa; open circles) and from ref.~\cite{collignon2020} (ambient; dots), in the $H \rightarrow 0$ limit obtained via back-extrapolations as shown in Fig.~\ref{SvsH}. Data for $\rho$ and $R_{\rm H}$ are reproduced from ref.~\cite{collignon2017}, and are respectively in zero-field (grey) and $H$ = 16~T at high temperature, and $H$~=~33~T at low temperature. In panel (b), we label the value of $\rho$ at $p = 0.22$ and $H$~=~33~T, extrapolated to $T \rightarrow 0$, as $\rho(0)$, and the value obtained from a linear extrapolation of the high temperature $T$-linear regime as $\rho_0$.
Bottom row (2.0~GPa): data on Nd-LSCO at $p = 0.22$ under a pressure of 2.0~GPa, showing a clear suppression of the low-temperature normal-state $S/T$ (d), $\rho$ (e), and $R_{\rm H}$ (f). Data for $S/T$ are from the present study in $H$ = 31~T. Data for $\rho$ and $R_{\rm H}$ are reproduced from ref.~\cite{doiron-leyraud2017}, in fields as indicated.
}
\label{All}
\end{figure*}
%

\section{Methods}

We measured the Seebeck coefficient using a low AC-technique derived from the Angstr\"om method~\cite{zhu2016,wang2019} and adapted specifically for experiments in the pressure transmitting fluid encapsulated in a piston-cylinder pressure cell.
Our experimental setup is displayed in inset of Fig.~\ref{Phase}.
A thermal excitation was applied by sending an electrical current through a 120~$\Omega$ strain gauge heater located on one side of the pressure chamber.
The resulting longitudinal thermal gradient $\Delta T_x$ accross the sample was measured using a type E differential thermocouple directly anchored to the sample.
An absolute type E thermocouple measures the temperature on the cold side of the differential thermocouple, $T^{-}$, allowing a determination of the average temperature of the sample $T_{av} = T^{-} + \Delta T_x/2$.
In all our measurements $T_{av}$ was found to be very close to $T_0$, the temperature of our Cernox sensor located just outside the pressure cell, meaning that the heat current generated a small $\Delta T_x$ without much elevation of the temperature inside the pressure cell.
The Seebeck voltage $\Delta V_x$ was measured with phosphor-bronze wires using the same contacts as $\Delta T_x$, which eliminates uncertainties associated with the geometric factor.
The Seebeck coefficient is then given by $S = - \Delta V_x / \Delta T_x$.
For the heat current we use a low AC square wave electrical current oscillating between zero and twice the wave amplitude.
The thermocouples and Seebeck voltages are amplified using preamplifiers based on EM Electronics A10 chips and picked-up using SR830 lock-in amplifiers at the thermal excitation frequency.
Our AC method provides two major advantages over the usual steady-state DC technique: 1) a major boost in measuring speed, which allows recording $S(T)$ continuously from 2 to 300K within a few hours and, 2) a greater stability against noise and perturbations that typically plague DC measurements.
We carefully benchmarked our approach against the DC method and found no significant difference.

Pressure was applied on our samples using a miniature non-magnetic piston-cylinder cell.
The pressure medium is Daphne oil 7474, which remains liquid at all pressures measured here at 300 K, ensuring a high degree of hydrostaticity.
The internal pressure is measured both at room temperature and at 4.2 K, using either the fluorescence of a small ruby chip or a Sn manometer.
The values quoted throughout are the low temperature pressures.
The error bar on all the pressure values is $\pm$~0.05 GPa, which comes from the uncertainty in measuring the position of the fluorescence peaks
For each measurement, the cell was cooled slowly ($<~$1 K/min) to ensure a homogeneous freezing of the pressure medium.

Large single crystals of Nd-LSCO were grown at Texas Materials Institute by a traveling float-zone technique in an image furnace, with nominal Sr concentrations $x = 0.12$, 0.22, and 0.24.
Two of these samples (0.22 and 0.24) were previously measured by electrical resistivity and Hall effect~\cite{collignon2017}, and all three were studied by thermal conductivity~\cite{michon2019} (sample details can be found in these references).
Our crystal of Eu-LSCO with $x = 0.125$ was grown in Tokyo.
Thermopower measurements on a closely related sample were previously reported in ref.~\cite{laliberte2011}, further sample details can be found there.
The hole concentration $p$ of each sample is given by $p = x$.
Samples were cut into small rectangular platelets of typical dimensions 1 mm $\times$ 0.5 mm $\times$ 0.2 mm, with the shortest dimension along the $c$ axis.
Contacts were made with H20E silver epoxy diffused by annealing at high temperature in flowing oxygen.
Thermopower measurements under pressure in magnetic fields up to 18~T were performed at Sherbrooke, and up to 31~T at the NHMFL in Tallahassee.
The magnetic field was applied along the $c$ axis and the Seebeck voltage signal was symmetrized with respect to field inversion in order to remove contaminations from the Nernst effect.

\section{Results and Discussion}

\subsection{Pseudogap phase at ambient pressure}
In Fig.~\ref{SvsH} we show isotherms of $S/T$ as a function of magnetic field $H$ up to 31~T for Nd-LSCO samples at $p = 0.22$ and 0.24 under pressure.
With increasing field, $S/T$ is null in the superconducting state at low field, then rises quickly upon crossing the vortex solid melting field $H_{\rm vs}$, and finally reaches the normal state value above the upper critical field $H_{c2}$.
At both dopings, we observe a rise of the normal state $S/T$ with decreasing temperature.
These findings are in overall agreement with our previous report of the zero-pressure thermopower in Nd-LSCO~\cite{collignon2020}.
In particular, as shown in Fig.~\ref{All}(a), our data inside a pressure cell at the lowest possible pressure of 0.1~GPa are in excellent quantitative agreement with ambient pressure data on the same sample, free-standing and measured using a standard steady-state DC method~\cite{collignon2020}, which demonstrates the reliability of our pressure setup.
This holds for both $p = 0.22$ and 0.24.
As in ref.~\cite{collignon2020}, these curves for $S/T$ versus $T$ as $H \rightarrow 0$ are constructed from the isotherms shown in Fig.~\ref{SvsH}, using the linear fits to the normal state data above $H_{c2}$, back-extrapolated to $H = 0$ in order to capture the intrinsic normal-state $S/T$ free from the sample-dependent negative magneto-Seebeck effect (which is similar to the magnetoresistance of normal metals).

\begin{figure}[t!]
\includegraphics[width=0.45\textwidth]{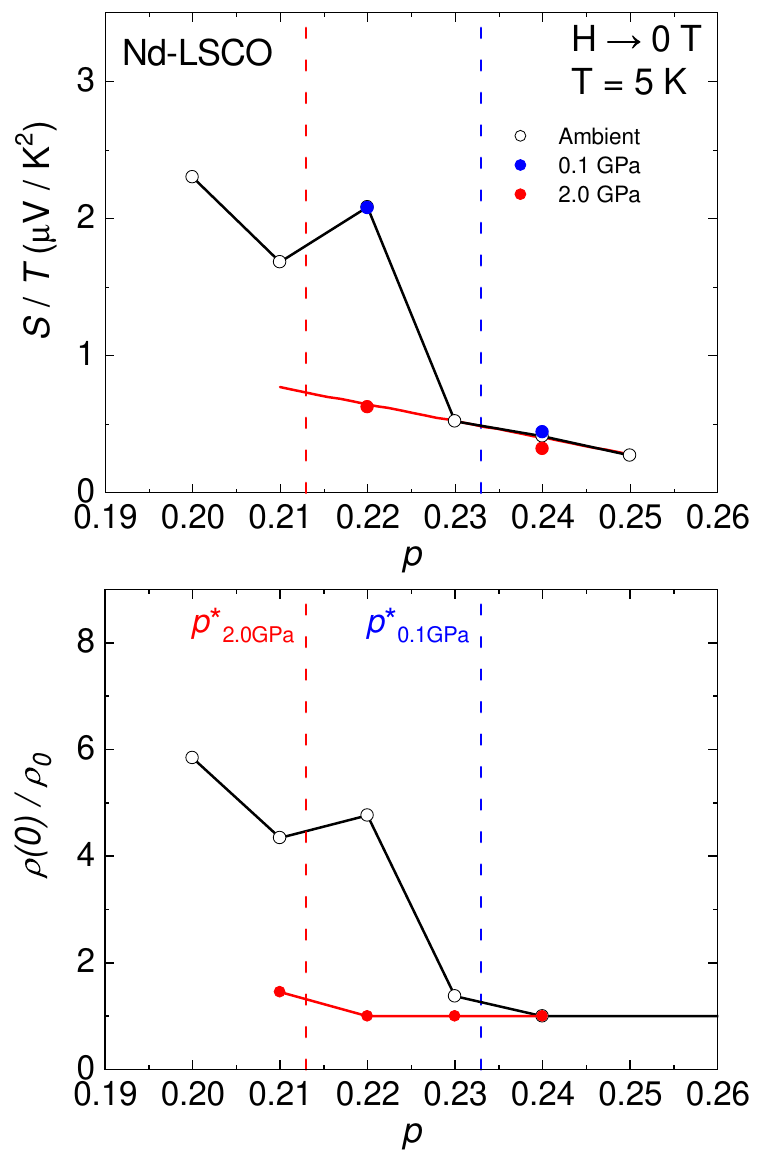}
\caption{
Top: Seebeck coefficient $S/T$ versus doping $p$ for Nd-LSCO, as measured at $T = 5$~K in the $H \rightarrow 0$ limit. Data at ambient pressure (open circles) are reproduced from~\cite{collignon2020}. Data at $P = 0.1$~GPa (blue dots) and 2.0~GPa (red dots) are from the present study. The vertical blue dashed line indicates $p^{\star}$ at ambient pressure, which coincides with the onset of the rise of $S/T$ caused by the drop in carrier density. Upon application of 2.0~GPa this rise in $S/T$ is fully suppressed at $p = 0.22$, and $S/T$ now follows the doping evolution extrapolated from outside the pseudogap phase (red line).
Bottom: Ratio $\rho(0)/\rho_0$ (see Fig.~\ref{All}(b)) as a function of doping, at ambient pressure (open circles~\cite{collignon2017}) and 2.0~GPa (red dots~\cite{doiron-leyraud2017}), which puts $p^{\star}$ in 2.0~GPa slightly above 0.21~\cite{doiron-leyraud2017}, as shown by the vertical red dashed line. The present data for $S/T$ are consistent with such a pressure-induced suppression of $p^{\star}$.
}
\label{Svsp}
\end{figure}

At $P$~=~0.1~GPa (Fig.~\ref{SvsH}), the curves of $S/T (H \rightarrow 0)$ yield, at 5~K, $S/T = 0.45$ and 2.1~$\mu$V / K$^2$ at $p = 0.24$ and 0.22, respectively, roughly a 5-fold increase.
As a function of doping, this increase in $S/T (H \rightarrow 0)$ suddenly occurs at $p^{\star}$, as shown in Fig.~\ref{Svsp} and reported in ref.~\cite{collignon2020}, and constitutes a clear thermopower signature of the pseudogap phase.
It confirms and reinforces the pseudogap signatures in the electrical resistivity $\rho$ and Hall coefficient $R_{\rm H}$~\cite{collignon2017}, as shown in the top row of Fig.~\ref{All} where we show the parallel low-temperature upturns in $S/T$, $\rho$, and $R_{\rm H}$ in Nd-LSCO $p = 0.22$.
As a result, in the $T \rightarrow 0$ limit the normal-state values of $\rho$ and $S/T$ display a comparable increase between $p = 0.24$ and 0.22, by a factor of about 5 (Fig.~\ref{Svsp}).
Since, in a simple model, $S/T \propto m^{\star}/n$, $\rho \propto m^{\star}/\tau n$, and $R_{\rm H} \propto 1/n$, where $1/\tau$ is the scattering rate, this strongly suggests that it is a drop in carrier density that causes the jump in the three quantities across $p^{\star}$.
As for YBCO~\cite{badoux2016} and Bi2201~\cite{lizaire2020}, these transport signatures in Nd-LSCO are all indicative of a drop in carrier density at $p^{\star}$, with $n$ going from $n = 1 + p$ above $p^{\star}$ to $n = p$ below.

Going just outside the pseudogap phase, at $p = 0.24$, Nd-LSCO instead displays signatures of quantum criticality, as first noted by the linear-$T$ resistivity which extends down to the lowest measured temperature~\cite{daou2009} (Fig.~\ref{All}(b)) and whose slope was recently shown to obey the Planckian limit~\cite{legros2019}, seen in several other quantum critical metals~\cite{bruin2013}. 
Recent specific heat measurements on Nd-LSCO revealed $C_{el}/T \propto$~Log(1/$T$) at $p = 0.24$~\cite{michon2019}, another signature of quantum criticality~\cite{lohneysen2007}. A logarithmic divergence is also seen in the Seebeck coefficient, with $S/T \propto$~Log(1/$T$) at $p = 0.24$ in both Nd-LSCO~\cite{daou_thermo2009} and Eu-LSCO~\cite{laliberte2011}.

%
%

\subsection{Pseudogap phase at 2.0~GPa}
Our main result is displayed in Fig.~\ref{All}(d): the low-temperature upturn in $S/T$ at $p = 0.22$ is fully suppressed under a pressure of 2.0~GPa, with $S/T$ now increasing only very slowly with decreasing temperature.
Quantitatively, $S/T$ at $p = 0.22$ at 5~K and 31~T goes from 1.8~$\mu$V / K$^2$ in 0.1~GPa to 0.7~$\mu$V / K$^2$ in 2.0~GPa.
As a result of this suppression, $S/T$ at $p = 0.22$ and 2.0~GPa displays essentially the same temperature evolution as that for $p = 0.24$ at 0.1~GPa, except for a rigid shift.
Note that we show continuous $T$-sweeps of $S/T$ in $H$ = 31~T taken using our AC method, something not possible with a standard DC technique.
Nevertheless, as displayed in Fig.~\ref{SvsH}(c), the profile of the raw field sweeps (taken with the AC method) of $S/T$ at $p = 0.22$ and 2.0~GPa are similar to those at 0.1~GPa, except for the strong reduction in the amplitude of $S/T$.
This pressure suppression of $S/T$ at $p = 0.22$ mirrors the suppression first observed in $\rho$ and $R_{\rm H}$~\cite{doiron-leyraud2017} and displayed for the same doping in Fig.~\ref{All}(e) and (f): in 2.0 GPa, $\rho$ goes from showing a huge low-temperature upturn at ambient pressure to displaying a pure linear-$T$ behavior, and $R_{\rm H}$ looses its upturn and becomes flat.
Having now the Seebeck coefficient showing the same suppression with pressure demonstrates that the effect is not a peculiarity of $\rho$ or $R_{\rm H}$, but is the reflection of a genuine suppression of the drop in carrier density accompanying the pseudogap phase, with all three coefficients at $p = 0.22$ and 2.0~GPa displaying the same behavior as at $p = 0.24$, where there is no pseudogap.
In contrast, we note that 2.0~GPa has a marginal effect on our $p = 0.24$ sample, slightly flattening $S/T$ at low temperature (Fig.~\ref{All}(d)).
Another effect of pressure is a strengthening of superconductivity at $p = 0.22$, as indicated by the rise of $T_c$ and $H_{c2}$ (Fig.~\ref{SvsH}).

We observe that pressure impacts only the low temperature behavior, with the curves of $S/T$, $\rho$, and $R_{\rm H}$ in 2.0~GPa all merging with the ambient pressure (or 0.1~GPa) curves at some temperature above 40~K or so (Figs.~\ref{All}(d,e,f)).
Pressure also does not change the actual doping in Nd-LSCO, which is determined by the Sr content.
In Fig.~\ref{Svsp} we show our data for $S/T$ at 5~K and $H \rightarrow 0$ at 2.0~GPa and observe that the point at $p = 0.22$ naturally extrapolates the line of $S/T$ vs $p$ above $p^{\star}$, showing that $p^{\star}$ itself has moved to lower dopings in 2.0~GPa.
Based on our resistivity data under pressure~\cite{doiron-leyraud2017}, shown in Fig.~\ref{Svsp}, we see that $p^{\star}$, as signaled by the sudden jump in $\rho(0)/\rho_0$, has in fact moved from 0.23 to about 0.21 in 2.0~GPa.
This implies a rate of suppression of $dp^{\star}/dP \simeq - 0.01$~hole per Cu atom/GPa, consistent with our Seebeck data at $p = 0.22$.
We expect that $S/T$ at $p = 0.21$ should also display a near full suppression under 2.0~GPa.
As discussed in ref.~\cite{doiron-leyraud2017}, we stress that pressure has no effect on the pseudogap temperature $T^{\star}$ itself, so the suppression of $p^{\star}$ is not accompanied by an overall collapse of $T^{\star}$, which is a huge energy scale.

Consistent with the fact that $p^{\star}$ moves down with pressure, we observe that $S/T$ at $p = 0.22$ and 0.24, which are both outside the pseudogap phase in 2.0~GPa, exhibit the same slow growth with decreasing temperature but with a roughly 2-fold difference in size (Fig.~\ref{All}(d)).
This is the same factor by which the inelastic part of the resistivity, which is perfectly $T$-linear, changes between $p = 0.22$ and 0.24 (Fig.~\ref{All}(b,e)).
Given that $S/T \propto m^{\star}/n$, $\rho \propto m^{\star}/\tau n$, and that $n$ varies little over this doping range (for $p > p^{\star}$), we infer that the factor of 2 must come from an increase of $m^{\star}$ upon approaching $p^{\star}$ from above, as argued in ref.~\cite{legros2019}. Such an increase of $m^{\star}$ is consistent with specific heat data on Nd-LSCO~\cite{michon2019} that show an increase of $C_{el}/T$ as $p$ is lowered towards $p^{\star}$ from above.
So the quantitative values of $S/T$ and $\rho$ in 2.0~GPa are consistent with the doping evolution expected at $p > p^{\star}$.

The fact that the shift of $p^{\star}$ under pressure is now seen in thermopower confirms and reinforces the conclusion laid out in ref.~\cite{doiron-leyraud2017}, namely that the shift of $p_{\rm FS}$ with pressure is driving a corresponding shift in $p^{\star}$, such that $p^{\star} \leq p_{\rm FS}$ continues to be obeyed.
Above $p_{\rm FS}$, the Fermi surface is electron-like and the Hall coefficient $R_{\rm H}$ is seen to linearly decrease with doping~\cite{ando2004,tsukada2006}, reaching negative values well above $p_{\rm FS}$.
Consequently, a relative change of $R_{\rm H}$ with pressure at constant doping indicates that $p_{\rm FS}$ itself is moving with pressure, as observed~\cite{doiron-leyraud2017} and expected from band-structure calculations~\cite{doiron-leyraud2017}.
The shift of $p_{\rm FS}$ and $p^{\star}$ in 2.0~GPa were found to match, both moving by about 0.02~holes per planar Cu atoms~\cite{doiron-leyraud2017}.
That $p^{\star} \leq p_{\rm FS}$ must be obeyed explains why LSCO, Nd/Eu-LSCO, and Bi2201 all have different $p^{\star}$, and shows that the pseudogap can only exist on a hole-like Fermi surface, which imposes a stringent constraint on theories of the pseudogap phase.
This was also found in numerical calculations of the Hubbard model~\cite{wu2018}.

%
\begin{figure}[t!]
\includegraphics[width=0.45\textwidth]{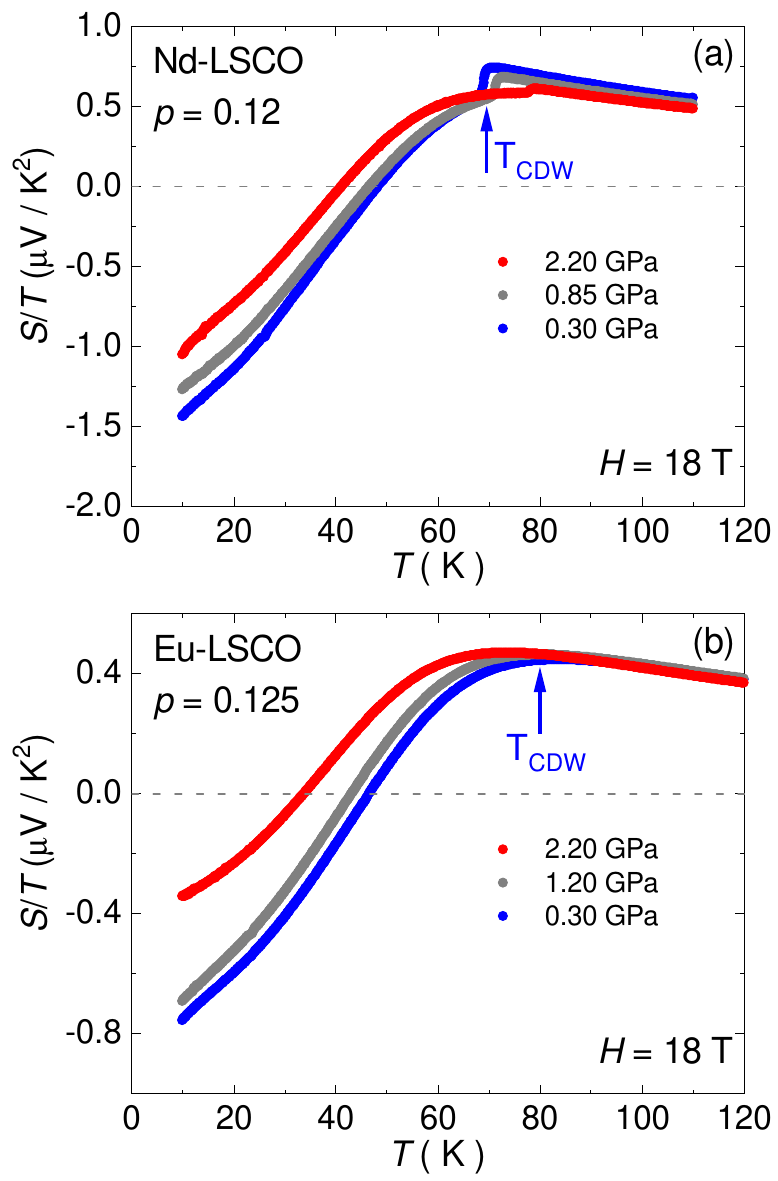}
\caption{Seebeck coefficient $S/T$ versus $T$ in the CDW phase, in (a) Nd-LSCO at $p = 0.12$ and (b) Eu-LSCO at $p = 0.125$, in the field-induced normal state at $H = 18$~T and at pressures as indicated. In Nd-LSCO $p = 0.12$, x-ray diffraction measurements~\cite{zimmermann1998,niemoller1999} detect the onset of CDW order at $T_{\rm CDW} = 70$~K, which coincides with the maximum in $S/T$ at $T_{\rm max} \simeq 70$~K. The step in the data is caused by the LTO-LTT structural transition. In Eu-LSCO 0.125, the CDW order occurs at $T_{\rm CDW} \simeq$~80~K~\cite{fink2011} while the structural transition is at $T_{\rm LTT} \simeq 130$~K~\cite{klauss2000}.
}
\label{0p12}
\end{figure}
%

\subsection{Pressure effect on charge density-wave phase}

We now turn to dopings well below $p^{\star}$ and at the center of the charge density-wave (CDW) phase, namely $p = 0.12$ in Nd-LSCO and $p = 0.125$ in Eu-LSCO.
As displayed in Fig.~\ref{Phase}, x-ray diffraction measurements~\cite{zimmermann1998,niemoller1999} detect the onset of CDW order in Nd-LSCO $p = 0.12$ at a temperature $T_{\rm CDW} = 70$~K.
Our data for $S/T$ versus $T$ for this sample show a clear departure from the data at $p = 0.22$, with $S/T$ at near ambient pressure (0.3~GPa) going through a broad maximum at $T_{\rm max} \simeq 70$~K before falling to negative values at low temperatures (Fig.~\ref{0p12}(a)).
Eu-LSCO at $p = 0.125$ exhibits a similar behavior for $S/T$ (Fig.~\ref{0p12}(b)), while x-ray measurements find a comparable $T_{\rm CDW}$ close to 80~K~\cite{fink2011}.
A negative $S/T$ is typical of cuprates near $p \simeq 1/8$, as observed in
LSCO~\cite{badoux2016a},
LBCO~\cite{Li2007}, 
Nd-LSCO~\cite{nakamura1992,hucker1998},
Eu-LSCO~\cite{hucker1998,chang2010}, 
YBCO~\cite{chang2010,laliberte2011}, and
Hg1201~\cite{doiron-leyraud2013}.
These studies showed that a negative Seebeck coefficient is a consequence of the Fermi surface reconstruction caused by the CDW order.
In both Nd-LSCO and Eu-LSCO, this is supported by the fact that $T_{\rm max}$ coincides with $T_{\rm CDW}$ (Fig.~\ref{Phase})~\cite{collignon2020}.
This recently allowed us to establish the doping range of the CDW phase in Nd-LSCO, present up to about $p =  p_{\rm CDW} \simeq 0.19$~\cite{collignon2020}.
Note that in Nd-LSCO $p = 0.12$, the structural transition from the low temperature orthorombic (LTO) to low temperature tetragonal (LTT) phase also coincides with $T_{\rm CDW}$ and $T_{\rm max}$, causing the sharp step in our $S/T$ data (Fig.~\ref{0p12}(a)).
In Eu-LSCO, the LTO-LTT transition occurs at a much higher temperature, with $T_{\rm LTT} \simeq 130$~K~\cite{klauss2000} at $p = 0.125$, and yet it displays the same $S/T$ curve as Nd-LSCO, showing that the CDW and not the LTO-LTT transition is causing the negative $S/T$~\cite{laliberte2011}.

In both Nd-LSCO and Eu-LSCO we observe two clear effects under increasing pressure: 1) at low temperatures, the amplitude of the negative $S/T$ is suppressed and 2) the temperature $T_{\rm max}$ shifts down.
In Eu-LSCO, the suppression of $S/T$ is significant, going at 10~K from about -0.8~$\mu$V/K$^2$ at 0.3~GPa to -0.35~$\mu$V/K$^2$ at 2.0~GPa, over a 50\% change.
$T_{\rm max}$ on the other hand goes from about 80~K at 0.3~GPa to 70~K at 2.0~GPa.
As a result, the sign-change temperature is also clearly suppressed by pressure, moving from 47~K at 0.3~GPa to 32~K at 2.0~GPa.
Similar effects are seen on Nd-LSCO, albeit smaller in amplitude.
Given that the negative $S/T$ is a clear marker of the CDW phase, its suppression is clear indication that the CDW phase is weakened by pressure.
Note that the high temperature $S/T$ above $T_{\rm max}$ is only weakly affected by the pressure, showing that pressure principally impacts the CDW phase.

Note also that superconductivity is boosted with pressure, as evidenced by the increase in $T_c$ we observe in both Nd-LSCO and Eu-LSCO (at $p \simeq 0.12$, \Tc~rises from 7~K at 0.3~GPa to 15~K at 2.2~GPa), suggesting a competition between SC and CDW.
This pressure tuning of the competition between SC and CDW was previously inferred in Nd-LSCO using transport measurements~\cite{arumugam2002} and in YBCO based on resistivity and Hall effect measurements under hydrostatic pressure~\cite{cyr-choiniere2018}.
In YBCO, the suppression of CDW modulations with pressure was directly observed by x-ray diffraction~\cite{souliou2018} and NMR~\cite{vinograd2019} measurements.
(Interestingly, in the case of Nd-LSCO 0.12 we observe a clear upward shift of $T_{\rm LTT}$ with pressure, in agreement with a previous x-ray study at the same doping~\cite{crawford2005}.)

\section{Summary}

We have used the Seebeck effect to examine the pressure dependence of the pseudogap critical point $p^{\star}$ in the cuprate superconductor Nd-LSCO, and of the CDW phase in both Nd-LSCO and Eu-LSCO.
We observe that the large Seebeck coefficient inside the pseudogap phase at $p = 0.22$, that results from the low carrier density below \pstar, is fully suppressed under a pressure $P$ = 2.0~GPa.
This confirms and reinforces our previous observation, inferred from resistivity and Hall effect measurements~\cite{doiron-leyraud2017}, that $p^{\star}$ in Nd-LSCO shifts down with pressure at the rate of $dp^{\star}/dP \simeq - 0.01$~hole per Cu atom/GPa and is fundamentally driven by a shift in pressure of the van Hove point where the Fermi surface changes from hole-like to electron-like.
This strengthens the notion that the pseudogap phase only exists on a hole-like Fermi surface, which implies important theoretical constraints.
At lower doping, at $p \simeq 1/8$ where the CDW phase is strongest, we observe in both Nd-LSCO and Eu-LSCO a reduction in magnitude of the negative Seebeck coefficient with increasing pressure, which we attribute to a weakening of the CDW order with pressure.
This confirms in Nd-LSCO and Eu-LSCO the phase competition between CDW order and superconductivity seen in other cuprates.

\section{Acknowledgements}

We thank S.~Fortier for his assistance with the experiments.
L.T. acknowledges support from the Canadian Institute for Advanced Research 
(CIFAR) as a CIFAR Fellow
and funding from 
the Institut Quantique, 
the Natural Sciences and Engineering Research Council of Canada (PIN:123817), 
the Fonds de Recherche du Qu\'ebec -- Nature et Technologies (FRQNT), 
the Canada Foundation for Innovation (CFI), 
and a Canada Research Chair.
This research was undertaken thanks in part to funding from the Canada First Research Excellence Fund
and the Gordon and Betty Moore Foundation's EPiQS Initiative (Grant GBMF5306 to L.T.).
The National High Magnetic Field Laboratory is supported by the National Science Foundation through NSF/DMR-1644779 and the State of Florida.
J.S.Z. was supported by NSF MRSEC under Cooperative Agreement No. DMR-1720595.



%

\end{document}